\documentstyle[11pt,epsf]{article}
\font\elevenbf=cmbx10 scaled\magstep 1
\font\elevenrm=cmr10 scaled\magstep 1
 1

\renewenvironment{thebibliography}[1]
 { \elevenrm
   \begin{list}{\arabic{enumi}.}
    {\usecounter{enumi} \setlength{\parsep}{0pt}
     \setlength{\itemsep}{3pt} \settowidth{\labelwidth}{#1.}
     \sloppy
    }}{\end{list}}
\renewcommand{\thefootnote}{\fnsymbol{footnote}}
\begin{document}
\noindent
\thispagestyle{empty}
\renewcommand{\thefootnote}{\fnsymbol{footnote}}
\begin{flushright}
{\bf TTP95-37}\footnote{The complete postscript file of this
preprint, including figures, is available via anonymous ftp at
ttpux2.physik.uni-karlsruhe.de (129.13.102.139) as
/ttp95-37/ttp95-37.ps or via www at
http://ttpux2.physik.uni-karlsruhe.de/cgi-bin/preprints/
Report-no: TTP95-37.}\\
{\bf October 1995}\\
hep-ph/9511393\\
\end{flushright}
\begin{center}
  \begin{large}
 RADIATION OF LIGHT FERMIONS IN HEAVY FERMION PRODUCTION\\
  \end{large}
  \vspace{0.5cm}
  \begin{large}
   A.H. Hoang\footnote{e--mail: hoang@ttpux1.physik.uni-karlsruhe.de},
   J.H. K\"uhn\footnote{Invited talk presented by J.H.~K\"uhn at
   the International
   Europhysics Conference, EPS-HEP, Brussels, July 27 -- August 2 1995.
   To appear in the proceedings.}
   and
   T. Teubner\footnote{e--mail: tt@ttpux2.physik.uni-karlsruhe.de}
 \\
  \end{large}
  \vspace{0.1cm}
Institut f\"ur Theoretische Teilchenphysik, Universit\"at Karlsruhe, \\
D--76128 Karlsruhe, Germany \\
  \vspace{0.7cm}
  {\bf Abstract}\\
\vspace{0.3cm}
\renewcommand{\thefootnote}{\arabic{footnote}}
\addtocounter{footnote}{-2}
\noindent
\begin{minipage}{15.0cm}
\begin{small}
Recent analytic calculations on
the rate for the production of a pair of massive fermions in $e^+ e^-$
annihilation plus real or virtual radiation of a pair of massless
fermions are discussed.
The contributions for real and virtual radiation are displayed separately.
The asymptotic behaviour close to threshold is given in a compact form and
an application to the angular distribution of massive quarks close to
threshold is presented.
\end{small}
\end{minipage}
\end{center}
\vspace{1.2cm}
%
%
\noindent
{\em 1.~Introduction}\\
The hadron production cross section belongs to the best measured quantities
in $e^+e^-$ collision experiments. Measurements exist in low and
intermediate energy ranges as well as around the $Z$ peak. A precise
perturbative determination of the heavy quark production cross
section $\sigma(e^+e^-\to\gamma^*\to Q\bar Q)$, ($Q=c,b,t$) is
therefore a crucial input to present and future interpretation of
experimental data. Especially future experiments at the $\tau$--charm
or the $B$ factory and the intended NLC require calculations valid
for the whole energy range above threshold. The
${\cal{O}}(\alpha_s)$ calculation including the full mass dependence
has been performed a long time
ago~\cite{schwinger}, whereas a complete second order computation
is still unknown. Asymptotic expansions~\cite{chetyrkin1}
of the second order
contribution are available in the limit of
vanishing fermion masses or including mass corrections
through an expansion in $m^2/s$ up to order $m^4/s^2$.
To third order the corresponding expansion is known up to order $m^2/s$.
These expansions, however, are inadequate near threshold.
\par
\begin{figure}[htb]
\begin{center}
\leavevmode
\epsfxsize=5cm
\epsffile[230 380 370 450]{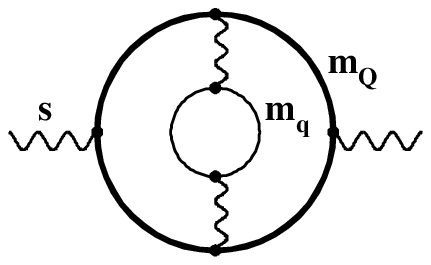}
\leavevmode
\epsfxsize=5cm
\epsffile[230 380 370 450]{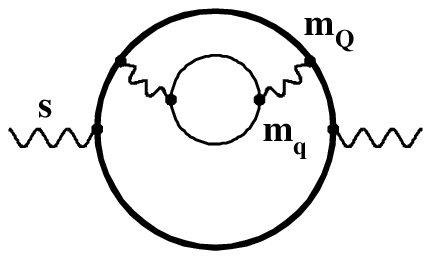}\\
\vskip 5mm
\caption{\label{fig1A} {\em Characteristic Feynman diagrams describing
the production of a pair of massive and a (real or virtual) pair of
massless fermions.}}
\end{center}
\end{figure}
\begin{figure}
\begin{center}
\leavevmode
\epsfxsize=8.5cm
\epsffile[120 300 450 530]{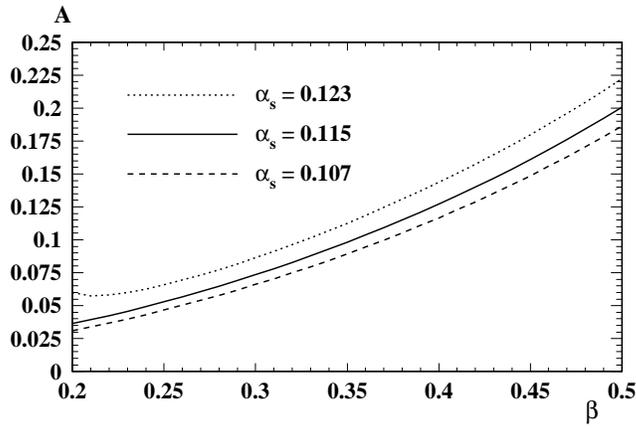}
\vskip -3mm
\caption[]{{Sensitivity of the anisotropy $A$ for
$e^+ e^- \to b\,\bar b$ to changes in
$\alpha_{\overline{MS}}(M^2_Z)$.}}
\label{fig3}
\end{center}
\end{figure}

This talk reports on recent analytic calculations of the complete
${\cal{O}}(\alpha_s^2)$ fermionic ($n_f$) contributions to the cross
section valid for the full physical energy range. This corresponds to
the evaluation of all possible cuts of the diagrams depicted in
Fig.~\ref{fig1A}. The resulting expressions describe real radiation of
a secondary fermion pair ($q\bar q$) off one of the primary
fermions ($Q\bar Q$) and the corresponding virtual effects.
\noindent
{\em 2.~Calculation}
The ${\cal{O}}(\alpha^2)$ QED calculation of the ratio
$R_{Q\bar Q q\bar q}=\sigma(e^+e^-\to\gamma^*\to Q\bar Q q\bar q)/\sigma_{pt}$,
where $\sigma_{pt}=4\pi\alpha^2/(3s)$, can be reduced to the following
phase space integral
\begin{equation}
R_{Q\bar Q q\bar q}=\frac{1}{3}\left(\frac{\alpha}{\pi}\right)
  \int\limits_{4m_q^2}^{(\sqrt{s}-2m_q)^2}\frac{\rm d\lambda^2}{\lambda^2}
  R_{Q\bar Q\gamma}(\lambda^2)R_{q\bar q}\,,
\label{masterreal}
\end{equation}
where $R_{Q\bar Q\gamma}$ and $R_{q\bar q}$ denote the the normalized
cross section of $Q\bar Q$ production with additional radiation of the
photon with mass $\lambda$ and of $q\bar q$ production respectively.
The corresponding virtual contribution reads
\begin{equation}
R_{Q\bar Q}^{q\bar q,\,virt}=\frac{1}{3}\left(\frac{\alpha}{\pi}\right)
\int\limits_{4m_q^2}^\infty\frac{\rm d\lambda^2}{\lambda^2}
R_{Q\bar Q}^{\gamma,\,virt}(\lambda^2)R_{q\bar q}\,,
\label{mastervirt}
\end{equation}
where $R_{Q\bar Q}^{\gamma,\,virt}$ describes the ${\cal{O}}(\alpha)$
virtual corrections to the $Q\bar Q$ production cross section due to
exchange of a massive photon with mass $\lambda$.
For the electromagnetic  Dirac ($F_{1}$) and Pauli ($F_{2}$)
form factors relations in close analogy to~(\ref{mastervirt})  exist.
The evaluation of eqs.~(\ref{masterreal}) and (\ref{mastervirt})
for $m_Q=0$ is fairly simple and was performed in~\cite{kniehl2,hoang1}.
The case of interest, $m_Q,s\gg m_q$, leads to much more involved
calculations~\cite{hoang2}.
The solutions have the following form
\begin{eqnarray}
R_{Q\bar Q q\bar q}=
\left(\frac{\alpha}{\pi}\right)^2
\left[ f_R^{(2)}\ln^2\frac{m_q^2}{s}+f_R^{(1)}\ln\frac{m_q^2}{s}+
f_R^{(0)} \right]\,, \mbox{\hspace{.3cm}}
\nonumber\\ \mbox{\hspace{.3cm}}
R_{Q\bar Q}^{q\bar q,\,virt}=
\left(\frac{\alpha_s}{\pi}\right)^2
\left[ f_V^{(2)}\ln^2\frac{m_q^2}{s}+f_V^{(1)}\ln\frac{m_q^2}{s}+
f_V^{(0)} \right]\,.\nonumber
\end{eqnarray}
Real and virtual corrections individually exhibit mass singularities
in the limit $m_q\to 0$. The quadratic logarithms correspond to
the soft photon singularities of the respective first order results.
However, $f_R^{(2)}=-f_V^{(2)}$, and thus at least the leading
logarithmic terms cancel if the sum of real and virtual
contributions is considered. A linear logarithm in $m_q^2/s$
remains in the sum and can be absorbed into a running
$\overline{\mbox{MS}}$ coupling $\alpha(\mu^2)$ at the scale $\mu^2=s$.
Due to the universality of the fermionic corrections in the second
order we only  have to replace one factor of $\alpha$ by
$2/3\,\alpha_s$ to arrive at the corresponding QCD expressions.
It is interesting that the functions
$f_{R/V}^{(1/2)}$ can already be deduced without any effort
from first order calculations.
Only $f_R^{(0)}$ and $f_V^{(0)}$ require an actual second order evaluation
and thus constitute the main effort. Their analytic results can
completely be expressed in terms of Tri- ($\rm{Li}_3$),
Di- ($\rm{Li}_2$) and usual logarithms.
{}From the complete analytic results high energy expansions in terms of
$m_Q^2/s$ can be derived which are in agreement with prior asymptotic
calculations~\cite{chetyrkin1}.
Near threshold the results
have to be expanded in terms of $\beta$, the relative velocity of the
$Q\bar Q$ pair. Real radiation is suppressed by $\beta^3$ whereas the
virtual contributions give a finite result at $\beta=0$.
Using the running coupling constant, defined by
\[ \alpha(\mu^2) = \alpha\Big[1-\frac{1}{3}\frac{\alpha}{\pi}
  \Big( \ln\frac{m_q^2}{\mu^2}+\frac{5}{3}\Big)\Big] \,,
\]
the combination $F_1+F_2$
which is relevant for the threshold cross section, assumes a particularly
simple form
\[
F_1+F_2=
1+\frac{\alpha(s\beta^2)\pi}{4\beta}-
2\frac{\alpha(se^{3/4}/4)}{\pi}+{\cal{O}}(\beta)\,.
\]
The scale in the
Coulomb exchange term $\propto 1/\beta$ is equal to the relative momentum of
the $Q\bar Q$ pair, whereas in the transvers photon term
the hard scale $\sim s/4$ has to be chosen. Up to second order in $\alpha$
the running coupling governing the Coulomb singularity is thus
identical to the coupling in the potential.
Similar conclusions can be drawn for each of the form factors individually.
Employing the BLM procedure this scale setting prescription can be
transfered to QCD.
\noindent
{\em 3.~Angular Distribution in $e^+e^-\to Q\bar Q$}
Because mesons with one heavy quark $Q$ ($Q=b,c$) follow the heavy quark
direction we can use our results to predict the meson angular
distribution near the $Q\bar Q$ production threshold\footnote{
\hspace{.2cm} Of course, the results can also be applied to $t\bar t$ or
$\tau^-\tau^+$ pair production near threshold.}. The quantity of interest
is the "anisotropy" $A$ in the cm angular distribution
\[
\frac{\rm d N}{\rm d\cos^2\theta}\sim
1+A\cos^2\theta \,,
\]
which measures the $D$--wave contribution with the characteristic
$\cos^2\theta$ behaviour relative to the $S$--wave. Due to the angular
distributions specific for the exclusive pseudoscalar or vector
meson channels the result can only be applied above the $B^*\bar{B}^*$ or
$D^*\bar{D}^*$ threshold, where the $Q\bar Q$ distribution is
saturated by the sum of the various channels ($B\bar{B}, B\bar{B}^*+cc.,
B^*\bar{B}^*$, etc.). To determine the anisotropy only virtual
corrections have to be taken into account. In the V-scheme this
leads to the result~\cite{brodsky}
\begin{eqnarray}
A=\tilde{A}/(1-\widetilde{A})\,,\mbox{\hspace{3cm}}\nonumber \\[2mm]
\widetilde A = \frac{\beta^2}{2}\,
 \frac{\left(1-4\,
\frac{\alpha_{\mbox{\it\tiny V}}(m^2 e^{7/6})}{\pi}\right)}
      {\left(1-\frac{16}{3}\,
      \frac{\alpha_{\mbox{\it\tiny V}}(m^2 e^{3/4})}{\pi}\right)}\,
 \frac{1-e^{-x_s}}{1-e^{-x_s^\prime}}\,
 \frac{\alpha_{\mbox{\it \tiny V}}(4\,m^2\,\beta^2/e)}
      {\alpha_{\mbox{\it \tiny V}}(4\,m^2\,\beta^2)}\,,
\nonumber
\end{eqnarray}
where
\[
x_s = \frac{4\,\pi}{3}\,
 \frac{\alpha_{\mbox{\it\scriptsize V}}(4\,m^2\beta^2)}{\beta}\,,
 \quad
x_s^\prime = \frac{4\,\pi}{3}\,\frac{\alpha_{\mbox{\it\scriptsize V}}
 (4\,m^2\beta^2/e)}{\beta}
\]
and a resummation of the Coulomb terms has been performed.
It is evident that in the anisotropy $A$ four different scales are involved.
In the factor arising from the Coulomb exchange the scale is smaller in the
numerator ($D$-wave) than in the denominator ($S$-wave). This behaviour is
consistent with qualitative considerations based on the relative distances
relevant for $S$- and $D$-waves in the Coulomb part.
In Fig.~\ref{fig3} the sensitivity of the anisotropy in the
$b\bar b$ production angular distribution is demonstrated for variations of
the value for $\alpha_{\overline{MS}}(M_Z^2)$.


\vskip 1.5cm
\sloppy
\raggedright
\def\app#1#2#3{{\it Act. Phys. Pol. }{\bf B #1} (#2) #3}
\def\apa#1#2#3{{\it Act. Phys. Austr.}{\bf #1} (#2) #3}
\def\lhc{Proc. LHC Workshop, CERN 90-10}
\def\npb#1#2#3{{\it Nucl. Phys. }{\bf B #1} (#2) #3}
\def\plb#1#2#3{{\it Phys. Lett. }{\bf B #1} (#2) #3}
\def\prd#1#2#3{{\it Phys. Rev. }{\bf D #1} (#2) #3}
\def\pR#1#2#3{{\it Phys. Rev. }{\bf #1} (#2) #3}
\def\prl#1#2#3{{\it Phys. Rev. Lett. }{\bf #1} (#2) #3}
\def\prc#1#2#3{{\it Phys. Reports }{\bf #1} (#2) #3}
\def\cpc#1#2#3{{\it Comp. Phys. Commun. }{\bf #1} (#2) #3}
\def\nim#1#2#3{{\it Nucl. Inst. Meth. }{\bf #1} (#2) #3}
\def\pr#1#2#3{{\it Phys. Reports }{\bf #1} (#2) #3}
\def\sovnp#1#2#3{{\it Sov. J. Nucl. Phys. }{\bf #1} (#2) #3}
\def\jl#1#2#3{{\it JETP Lett. }{\bf #1} (#2) #3}
\def\jet#1#2#3{{\it JETP Lett. }{\bf #1} (#2) #3}
\def\zpc#1#2#3{{\it Z. Phys. }{\bf C #1} (#2) #3}
\def\ptp#1#2#3{{\it Prog.~Theor.~Phys.~}{\bf #1} (#2) #3}
\def\nca#1#2#3{{\it Nouvo~Cim.~}{\bf #1A} (#2) #3}
{\elevenbf\noindent  References \hfil}
\vglue 0.4cm

\end{document}